\documentclass[aps,prd,twocolumn,amsmath,showpacs,superscriptaddress,nofootinbib]{revtex4-1}

\usepackage{graphicx}	
\usepackage{amsmath}	
\usepackage{amssymb}	

\def\be{\begin{equation}}
\def\ee{\end{equation}}

\def\bea{\begin{eqnarray}}
\def\eea{\end{eqnarray}}

\begin{document}
\title[Updating constraints on $f(T)$ teleparallel cosmology]{Updating constraints on $f(T)$ teleparallel cosmology and the consistency with Big Bang Nucleosynthesis}

\author{Micol Benetti}
\email{benettim@na.infn.it}
\affiliation{Dipartimento di Fisica  ``E. Pancini", Universit\`a di Napoli  ``Federico II", Via Cinthia, I-80126, Napoli, Italy}
\affiliation{Istituto Nazionale di Fisica Nucleare (INFN), sez. di Napoli, Via Cinthia 9, I-80126 Napoli, Italy}

\author{Salvatore Capozziello}
\affiliation{Dipartimento di Fisica  ``E. Pancini", Universit\`a di Napoli  ``Federico II", Via Cinthia, I-80126, Napoli, Italy}
\affiliation{Istituto Nazionale di Fisica Nucleare (INFN), sez. di Napoli, Via Cinthia 9, I-80126 Napoli, Italy}
\affiliation{Laboratory for Theoretical Cosmology, Tomsk State University of Control Systems and Radioelectronics (TUSUR), 634050 Tomsk,  Russia}

\author{Gaetano Lambiase}
\affiliation{Dipartimento di Fisica E.R. Cainaiello, University of Salerno, Via
Giovanni Paolo II, I 84084-Fisciano (SA), Italy}
\affiliation{INFN, Gruppo Collegato di Salerno, Sezione di Napoli, Via Giovanni Paolo II, I 84084-Fisciano (SA), Italy}

\begin{abstract}
We focus on viable $f(T)$ teleparallel cosmological  models, namely  power law,  exponential and  square-root exponential, carrying out a detailed study of their evolution at all scales. Indeed, these models were extensively analysed in the light of late time measurements, while it is possible to find only upper limits looking at the very early time behavior, i.e. satisfying the Big Bang Nucleosynthesis (BBN) data on primordial abundance of ${}^4He$. Starting from these indications,  we perform our analysis considering both background and linear perturbations evolution and constrain, beyond the standard six cosmological parameters, the free parameters of  $f(T)$ models in both cases whether the BBN consistency relation is considered or not.
We use a combination of Cosmic Microwave Background, Baryon Acoustic Oscillation, Supernovae Ia and galaxy clustering measurements, and find that  very narrow constraints on the free parameters of specific $f(T)$ cosmology can be obtained, beyond any previous precision. While no degeneration is found between the helium fraction, $Y_P$, and the free parameter of $f(T)$, we note that these models constrain the current Hubble parameter, $H_0$, {higher extent than  the standard model} one, fully compatible with the Riess et al. measurement in the case of power law $f(T)$ model. Moreover, the free parameters are constrained at non-zero values in more than 3-$\sigma$, showing a preference of the observations for extended gravity models.

\end{abstract}

\maketitle

\section{Introduction}

The Standard Cosmological Model, the so called $\Lambda$CDM, provides a reliable description of the Universe from some seconds after the big bang until the present epoch, under the assumptions that gravity is described by Einstein's General Relativity (GR), the spatial sections of the Universe, at constant cosmological time, are homogeneous and isotropic, and dark matter and dark energy components exist.
However, we know that the $\Lambda$CDM model is incomplete. For example, there is no final evidence  of dark matter and dark energy, nor explication for matter-antimatter asymmetry or unification of gravity and the other  interactions at quantum level (see Refs \cite{Capozziello:2014yqa,Altschul:2014lua,Tino:2020nla,Nojiri:2017ncd,Chakravarty:2016set,Lambiase:2013haa,Capolupo:2020agb,Buoninfante:2019uwo,Buoninfante:2018xiw,Buoninfante:2020ctr} and references therein).  Also, new physics beyond the Standard Model has been invoked to describe the increasingly precise data of the latest generation, since several tensions have emerged between data at different scales (for a detailed discussion see Refs. \cite{Verde:2019ivm,Benetti:2019lxu,Graef:2018fzu,Benetti:2017juy,Bernal:2016gxb,Guo:2018ans,Vattis:2019efj,Pan:2019gop} and references therein).

In this context, several assumptions have been re-considered, including the possibility of modifications and  extensions of  GR  in order to fix the dark energy and dark matter issues  due to  lack of evidences of these elements on a fundamental level. 
The paradigm of considering different  theories of gravity, with respect to GR, comes from the fact that Einstein's theory is proved to be not sufficient to describe dynamics of gravitational field at ultraviolet and infrared scales. According to this statement, several effective models have been proposed towards quantum gravity and cosmology with the aim to recover the agreement with the experiments and observations  reached by GR but enlarging also the number of phenomena to be described at different scales and energies \cite{Capozziello:2011et}.
The debate is not only related to the possibility of adding new contributions to the Hilbert-Einstein action, like in the case of $f(R)$ gravity and analogue theories, but also to identify the correct variables describing the gravitational field.  Equivalence Principle is assumed as the foundation of General Relativity and of  several metric theories \cite{Capozziello:2014yqa}. This assumption leads to the coincidence of the geodesic and causal structure and fixes the connection which as to be Levi-Civita.

Nevertheless Einstein himself recognized that such an approach could be enlarged and improved if alternative descriptions of gravitational dynamics were considered. In particular, if tetrads describe the gravitational field, dynamics can be given by torsion. In this picture, the Equivalence Principle is not the foundation of gravitational field and affinities assumes a fundamental role. These considerations led to the teleparallel formulation of GR which, at field equations level, is equivalent to GR giving the so called Teleparallel Equivalent General Relativity (TEGR). In this perspective, also extensions of TEGR reveal interesting and then, as the straightforward extension of curvature gravity is $f(R)$ (where $R$ is the Ricci scalar), now $f(T)$ extends TEGR (being $T$ the torsion scalar).

One of the main goals to develop these alternative approach is to select self-consistent cosmological models capable of giving a realistic picture of cosmic history (see \cite{Cai:2015emx,Barker:2020elg,Barker:2020gcp} for a detailed discussion). The goal is to coherently connect early (inflation) and late Universe (dark energy), passing for large scale structure formation. In this program cosmography \cite{Capozziello:2020ctn,Capozziello:2019cav} and Big Bang Nucleosynthesis (BBN) \cite{Capozziello:2017bxm} could play a main role.
In particular, BBN offers one of the most powerful methods to test the validity of  cosmological models around the MeV energy scale.
The precise measure of the chemical abundances of the primordial elements of BBN is one of the main efforts of the modern cosmology \cite{Steigman:2007xt,Cyburt:2015mya,Mathews:2017xht,Steigman:2012ve}. Indeed, such  abundances of hydrogen, helium, lithium and deuterium are an important test for any  cosmological model, being extremely  sensitive to the physics of the early Universe.
Also, direct astrophysical observations allow to extrapolate primordial abundance. By the emission lines of nearby H$_{II}$ regions in metal-poor star forming galaxies,   the mass fraction of $^{4}$He ($Y_P$) has been sensitively estimated\cite{Izotov:2014fga,Aver:2015iza}, while the primordial $^{7}$Li abundance is determined by the atmospheres of very metal-poor stars \cite{Sbordone:2010zi,Francois:2013bba}. Finally, the primordial deuterium abundance can be measured using the absorption line of gas clouds \cite{Cooke:2013cba,Cooke:2016rky,Balashev:2015hoe,Riemer-Sorensen:2017pey,Zavarygin:2018ara}. Such a measurements allow a high precision estimate of the  baryon fraction density, and has been found a concordance between the Aver(2015) analysis \cite{Aver:2015iza} and the Planck(2018) derived ones \cite{Aghanim:2018eyx}. However,   also several tensions emerged, and they are  quantified in more then $2\sigma$, when $\Omega_b$ is derived by different model assumption of Ref.\cite{Izotov:2014fga} or deuterium abundance \cite{Cooke:2017cwo} (see also Ref.\cite{Aghanim:2018eyx} for an updated discussion of current results).

Although efforts are spent  to reconcile these measurements \cite{Cooke:2016rky,Cooke:2017cwo}, other possible cosmological models can be explored to test if a natural  agreement  can be obtained between the  $\Omega_b$ value inferred from the BBN and the derived one  from the Cosmic Microwave Background (CMB). For example, it is possible to bring closer the BBN and CMB predictions of the baryon density today considering  extensions to the Standard Model, such as a change in the expansion rate, parameterized by the effective number of relativistic degrees of freedom, $N_{eff}$ \cite{Abazajian:2013oma,Guo:2018ans,Benetti:2019lxu,Graef:2018fzu}.
Since both helium abundance and $N_{eff}$ affect the CMB damping tail, they are partially degenerate.
On the other hand, a phenomenological modeling of the current observed accelerated expansion of the Universe should  be  ideally  embedded  into  a  more  fundamental  framework, i.e. deduced from first principles. It is therefore timely to test fundamental theories with a study involving all scales, from the first seconds of the Universe (i.e. using BBN) to today observed accelerated expansion.

In this work, we focus on teleparallel gravity \cite{Cai:2015emx} and trace the observational prediction of different forms $f(T)$ using a Boltzmann numerical resolution code. 

Previous studies, analysing the high temperatures characterizing the primordial Universe, constrained with upper bounds the $f(T)$ cosmology \cite{Capozziello:2017bxm}, and it is timely to improve such an analysis using the wide range of available data at all scales. In this perspective, the feasibility of a teleparallel description of gravity can be realistically tested. In fact, until now, most efforts have been devoted to match late accelerated behavior by $f(T)$ gravity but the attempt to reproduce the whole cosmic history in a teleparallel picture has to be more pursued in order to finally compare metric and tetrad descriptions. Here, in particular,  we explore whether by relaxing the consistency of the BBN, it is possible that these theories are in agreement with the estimates of primordial abundances. This can be an important consistency test.\\

The  paper  is  organised  as  follows.
In  Section  \ref{Sec:Theory},  we  introduce    TEGR  and its $f(T)$ extension.  We will derive the related background cosmology   and the evolution of  primordial perturbations which  we will use for our analysis.
In  Section \ref{Sec:fT_models},  we  provide an overview of the specific models  we are going to analyse, showing observational predictions and giving a state of the art of current analyses.
Details of the  analysis method are reported in Section \ref{Sec:Analysis}, also indicating the data set we use to constrain the models parameters. Finally, in Section \ref{Sec:Results},  we discuss the  results and
draw our conclusions.

\section{$f(T)$ gravity and cosmology}
\label{Sec:Theory}

Let us  briefly review the main features of TEGR and  $f(T)$ teleparallel gravity. First, we introduce the vierbein fields
$e_i(x^\mu)$, $i = 0, 1, 2, 3$. They forms an orthonormal basis in the tangent space at each point $x^\mu$ of the manifold, i.e. $e_i \cdot e_j=\eta_{ij}$, with $\eta_{ij}=diag(1,-1,-1,-1)$ the Minkowski metric.
Denoting with $e^\mu_i$, with $\mu=0,1,2,3$ the components of the vectors $e_i$ in a
coordinate basis $\partial_\mu$, one can write $e_i=e^\mu_i\partial_\mu$ (the Latin indices refer to the tangent space, the Greek indices to the coordinates on
the manifold). The components of the metric tensor of the manifold, $g_{\mu\nu}(x)$ are constructed via the dual vierbein fields, i.e.
$g_{\mu\nu}(x)=\eta_{ij} e^i_\mu(x)e^j_\nu(x)$.

The TEGR models are characterized by the fact that
the curvatureless Weitzenb\"{o}ck connection is adopted (let us recall that, in General Relativity, one uses the torsion-less Levi-Civita
connection). This allows to define the non-null torsion tensor
\begin{equation}\label{torsion}
T^\lambda_{\mu\nu}=\hat{\Gamma}^\lambda_{\nu\mu}-\hat{\Gamma}^\lambda_{\mu\nu}
=e^\lambda_i(\partial_\mu e^i_\nu - \partial_\nu e^i_\mu).
\end{equation}
The action we are going to consider is of the form
\begin{equation}\label{action}
    I = \frac{1}{16\pi G}\int{d^4xe\left[T+f(T)\right]} + I_m,
\end{equation}
where $f(T)$ is a generic function of the torsion scalar $T$, $I_m$ is the action of matter fields, and $e=det(e^i_\mu)=\sqrt{-g}$ is the metric determinant. Explicitly,
the torsion scalar $T$ reads
\begin{equation}\label{lagrangian}
    T={S_\rho}^{\mu\nu}{T^\rho}_{\mu\nu}\,.
\end{equation}
\begin{eqnarray}
    {S_\rho}^{\mu\nu}&=&\frac{1}{2}({K^{\mu\nu}}_\rho+\delta^\mu_\rho
{T^{\theta\nu}}_\theta-\delta^\nu_\rho {T^{\theta\mu}}_\theta) \label{s} \\
    {K^{\mu\nu}}_\rho &=&
-\frac{1}{2}({T^{\mu\nu}}_\rho-{T^{\nu\mu}}_\rho-{T_\rho}^{\mu\nu})\,,
\label{contorsion}
\end{eqnarray}
with ${K^{\mu\nu}}_\rho$ the contorsion tensor which gives the difference between
Weitzenb\"{o}ck and Levi-Civita connections.

The variation with respect to the vierbein gives the field equations
\cite{Cai:2015emx}
\begin{multline}
 e^{-1}\partial_\mu(e e_i^\rho {S_\rho}^{\mu\nu})[1+f']-
   e_i^\lambda {T^\rho}_{\mu\lambda}{S_\rho}^{\nu\mu}[1+f']
  +e^\rho_i {S_\rho}^{\,\,\mu\nu}(\partial_\mu T)f'' + \\
      \frac{1}{4}e^\nu_i [T+f]=4\pi G\,{e_i}^\rho\, {\Theta_\rho}^\nu\,,
\end{multline}
where we defined  $f'\equiv df/dT$ and ${S_i}^{\mu\nu}={e_i}^\rho {S_\rho}^{\mu\nu}$, while
$\Theta_{\mu\nu}$ is the energy-momentum tensor of perfect fluid matter.

For a flat Friedmann-Lema\^itre-Robertson-Walker (FLRW) background, the metric is
\begin{equation}
ds^2= dt^2-a^2(t)\,\delta_{ij} dx^i dx^j,
\end{equation}
where $a(t)$ is the scale factor. The corresponding vierbien fields are $e_{\mu}^a={\rm diag}(1,a,a,a)$. The latter and
Eq. (\ref{lagrangian}) yield the relation between the torsion $T$ and
the Hubble parameter $T=-6H^2$, where ${\displaystyle H=\frac{\dot a}{a}}$.
Assuming that matter sector is described by a perfect fluid with energy density $\rho$ and pressure $p$, the field equations give 
\begin{equation}\label{friedmann}
    12H^2[1+f']+[T+f]=16\pi G\rho,
\end{equation}
\begin{equation}\label{acceleration}
    48H^2f''\dot{H}-(1+f')[12H^2+4\dot{H}]-(T-f)=16\pi Gp.
\end{equation}
Moreover, the equations are closed with the equation of continuity for the matter sector
$\dot{\rho}+3H(\rho+p)=0$.
Eqs. (\ref{friedmann}) and (\ref{acceleration}) can be rewritten in terms of
the effective energy density $\rho_T$ and pressure $p_T$ arising from $f(T)$
\begin{equation}\label{modfri}
    H^2=\frac{8\pi G}{3}(\rho+\rho_T),
\end{equation}
\begin{equation}\label{modacce}
    2\dot H+3H^2=-\frac{8\pi G}{3}(p+p_T)\,,
\end{equation}
where
\begin{eqnarray}\label{rhoT}
    \rho_T &=& \frac{3}{8\pi G}\left[\frac{Tf'}{3}-\frac{f}{6}\right],\\
    p_T&=&\frac{1}{16\pi G}\, \frac{f-T f'+ 2T^2f''}{1+f'+ 2{T}f''} \,,
\label{pT}
\end{eqnarray}
and define the effective torsion  equation-of-state
\begin{equation}\label{eq:omegaeff}
    \omega_{T}\equiv \frac{p_T}{\rho_T}= -\frac{f-T f'+2{T}^2f''}{(1+f'+ 2 T f'')(f-2Tf')}\,.
\end{equation}
These effective models are hence responsible for the accelerated phases of the early or/and late Universe \cite{Cai:2015emx}.

In order to perform our analysis, we rewrite the first FLRW equation, Eq.(\ref{modfri}), making explicit the form of the torsional energy density \cite{Nesseris:2013jea,Nunes:2018xbm,Nunes:2016qyp}

\begin{equation}
\label{eq:Ha}
\frac{H(a)^2}{H_0} \equiv E(a)^2 = \left[\Omega_{m0} a^{-3}+\Omega_{r0} a^{-4}+ \frac{1}{T_0} [f-2Tf'] \right] \,
\end{equation}

where we define $\Omega_{i0} = \frac{8 \pi G \rho_{i0}}{3H_0^2}$, and consider the relation $T=-6H^2$.
The above background evolution recovers the standard model  for $\frac{1}{T_0} [f-2Tf'] \rightarrow \Omega_{\Lambda}$. Hereafter we define such a torsional contribution as
\begin{equation}
 y_T(a,\xi) \equiv   \frac{1}{T_0} [f-2Tf']\,,
\end{equation}
with $\xi$ the free parameters of the $f(T)$ parameterization.

At this point, it is worth stressing that the main focus of the paper  is to analyze the $f(T)$ models at the level of background evolution according to the current expansion of the Universe. For our purpose, we consider an extended dataset, including both the BBN and the CMB as well as large-scale data.  In this perspective, we do not discuss in detail the theory of perturbations in $f(T)$ gravity (see, e.g.,  \cite{Golovnev:2018wbh}) but assume a standard primordial perturbation pattern for a perfect fluid, i.e. the ``torsion fluid" that drives the current expansion of the Universe via Eq.\eqref{eq:Ha}. Such a fluid  contributes  with its EoS defined in  Eqs.\eqref{eq:omegaeff}. In other words, the exact perturbative evolution of  specific $f (T)$ models  is beyond our interests and then we assume
a density contrast and velocity divergence in the synchronous gauge fixing  the torsion fluid frame with zero acceleration.  Specifically, we consider an effective  Bardeen approach where torsion contributes as   a zero-acceleration perfect fluid (\cite{Ma:1995ey,Hu:1998kj,Xu:2013mqe,Kumar:2012gr}). Furthermore, to support our choice, we will show  that the detailed treatment of Ref. \cite{Nunes:2018xbm}, adopting the power-law model, leads to the same result that we will obtain in the framework of our approximation.

In the next section, it is shown that the EoS of the considered models is close to that of  cosmological constant, and our assumption on perturbative sector does not significantly affect the observational predictions.

\begin{figure*}[t!]
\centering
\includegraphics[width=0.45\textwidth]{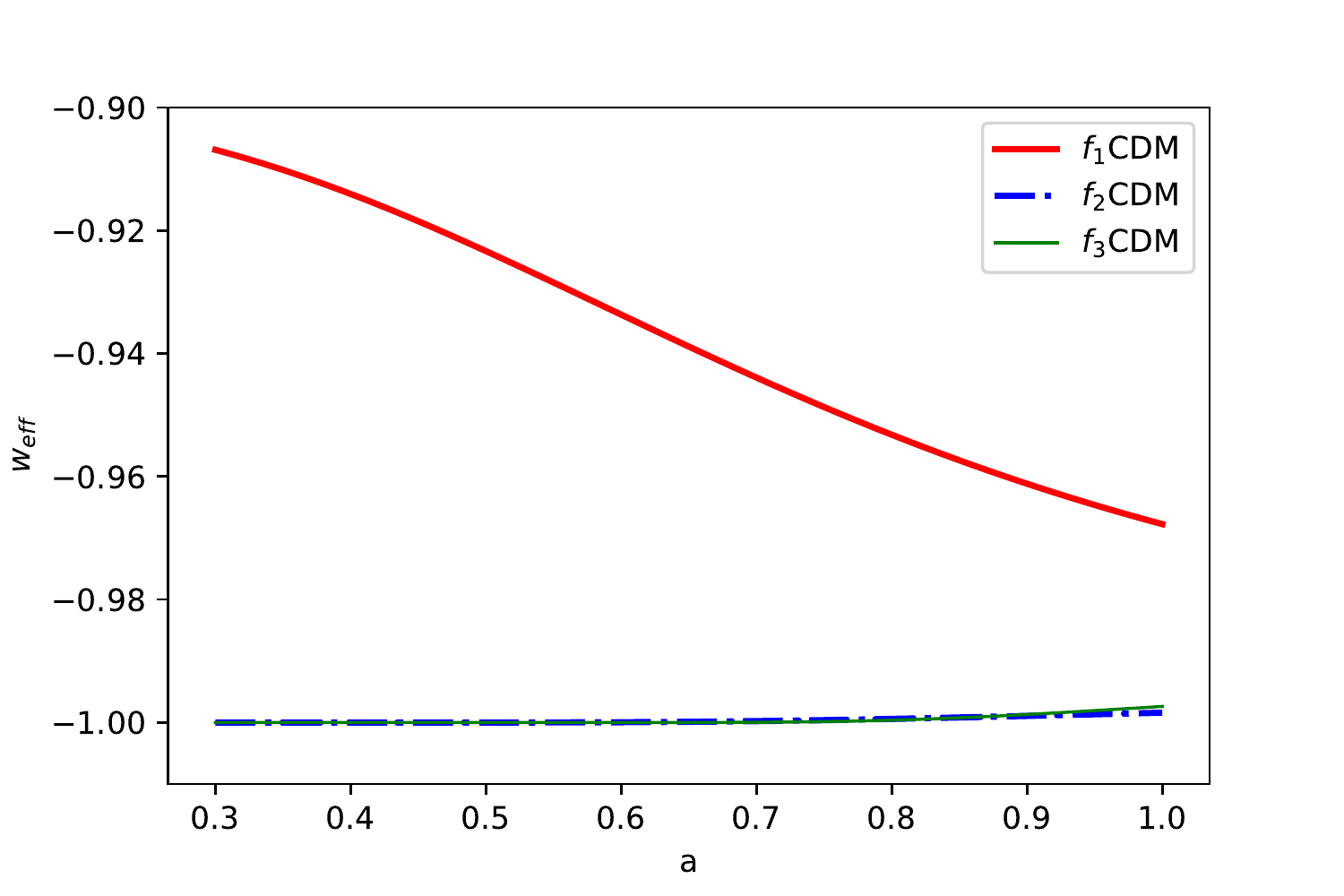}
\includegraphics[width=0.45\textwidth]{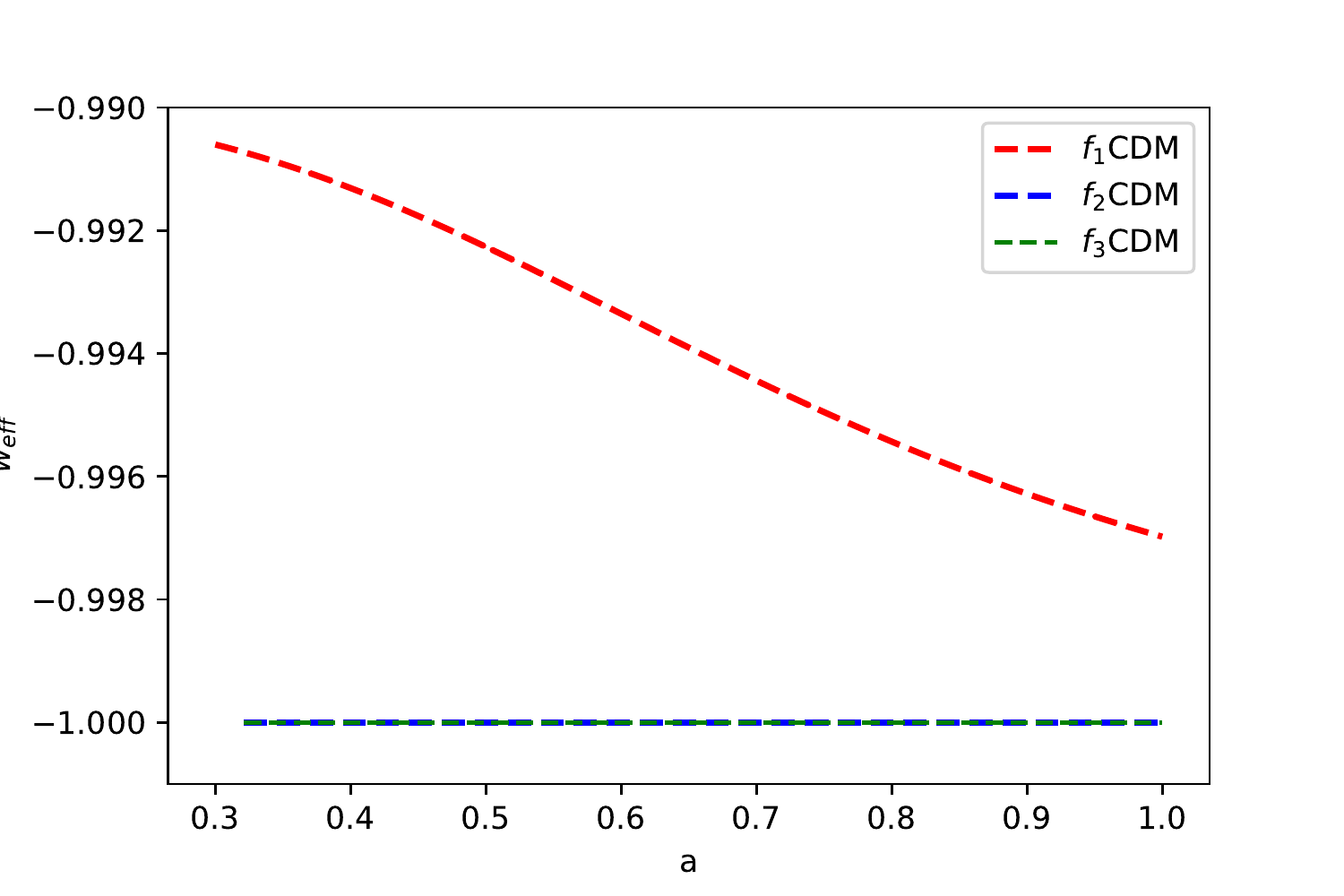}
\includegraphics[width=0.45\textwidth]{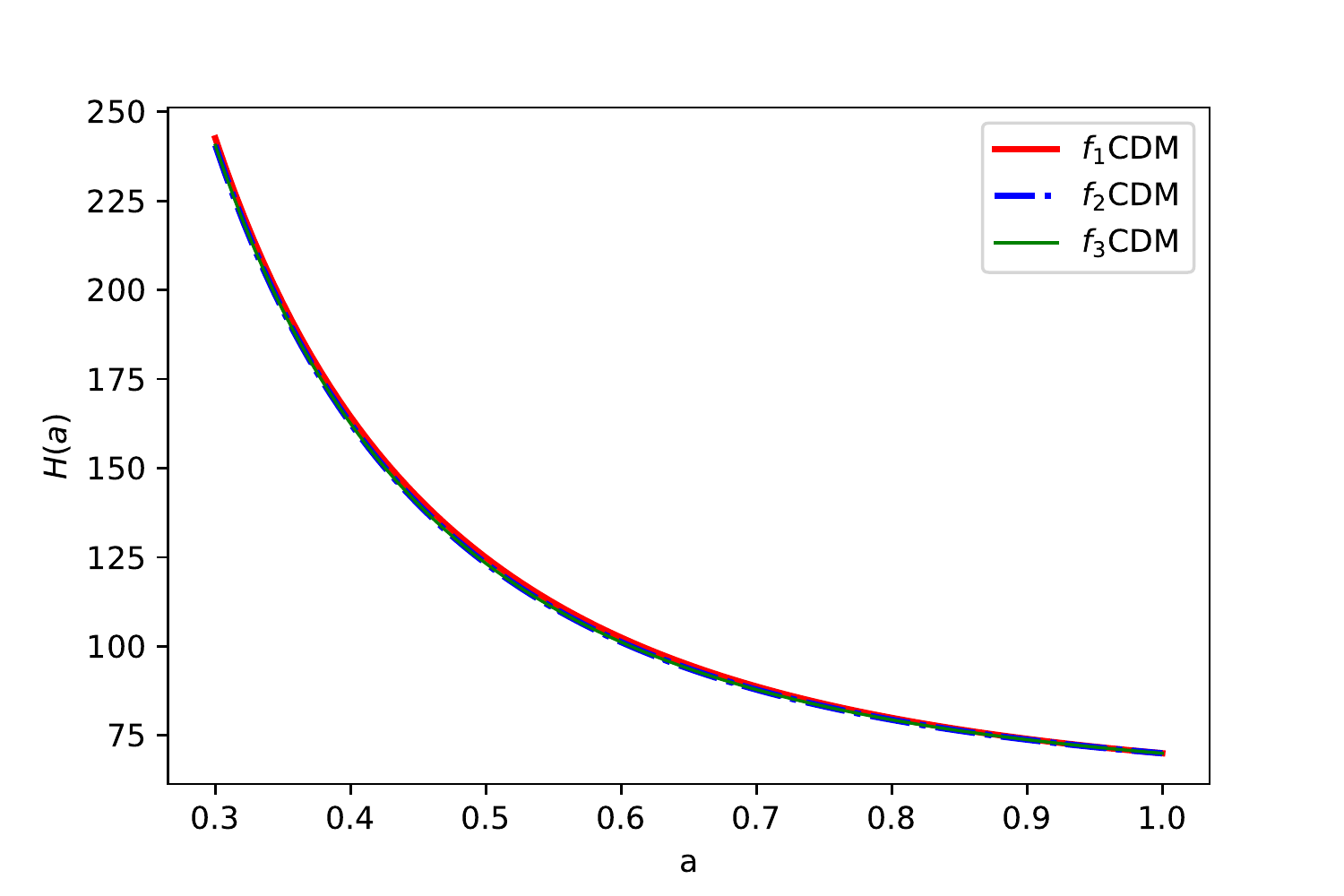}
\includegraphics[width=0.45\textwidth]{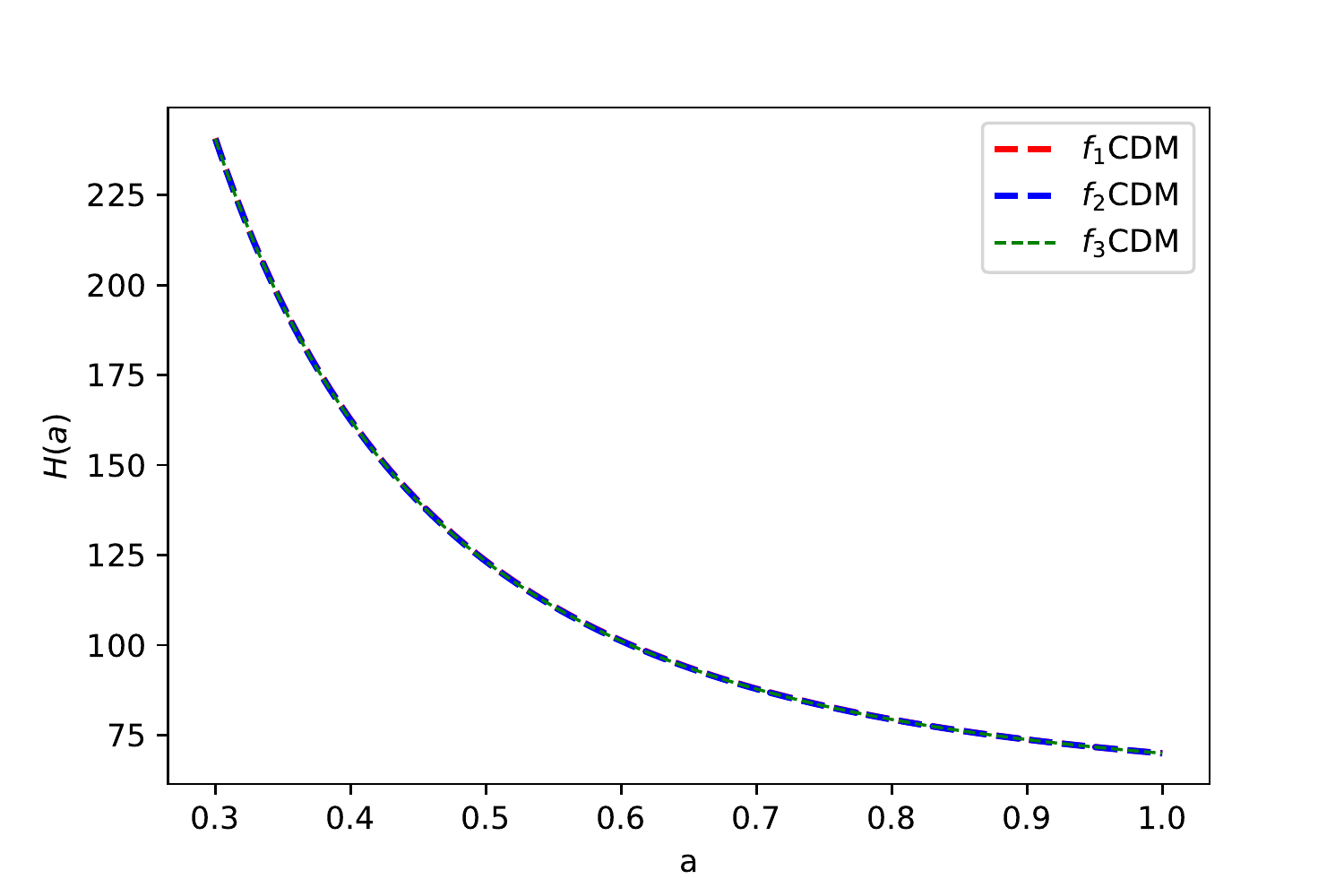}
\includegraphics[width=0.45\textwidth]{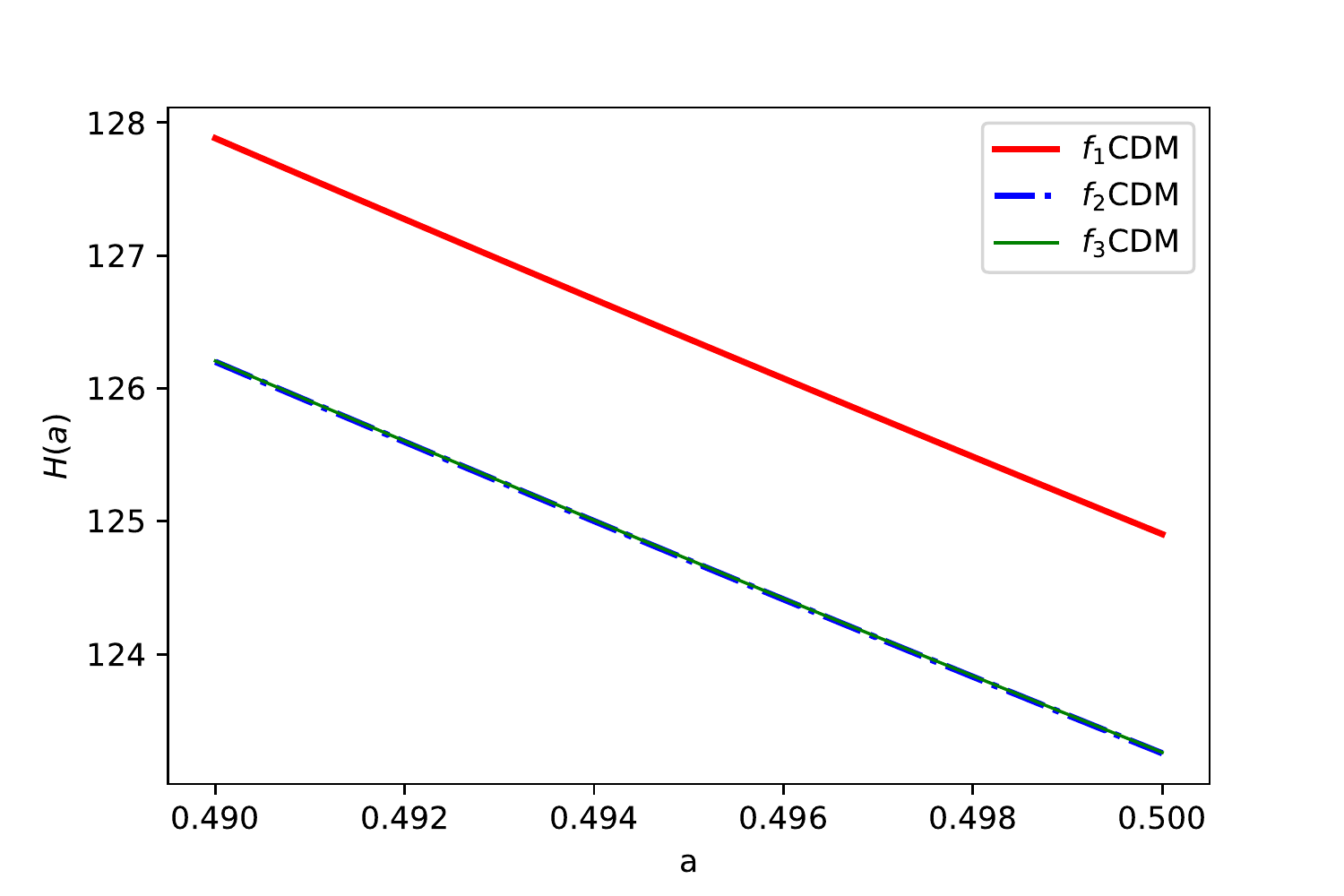}
\includegraphics[width=0.45\textwidth]{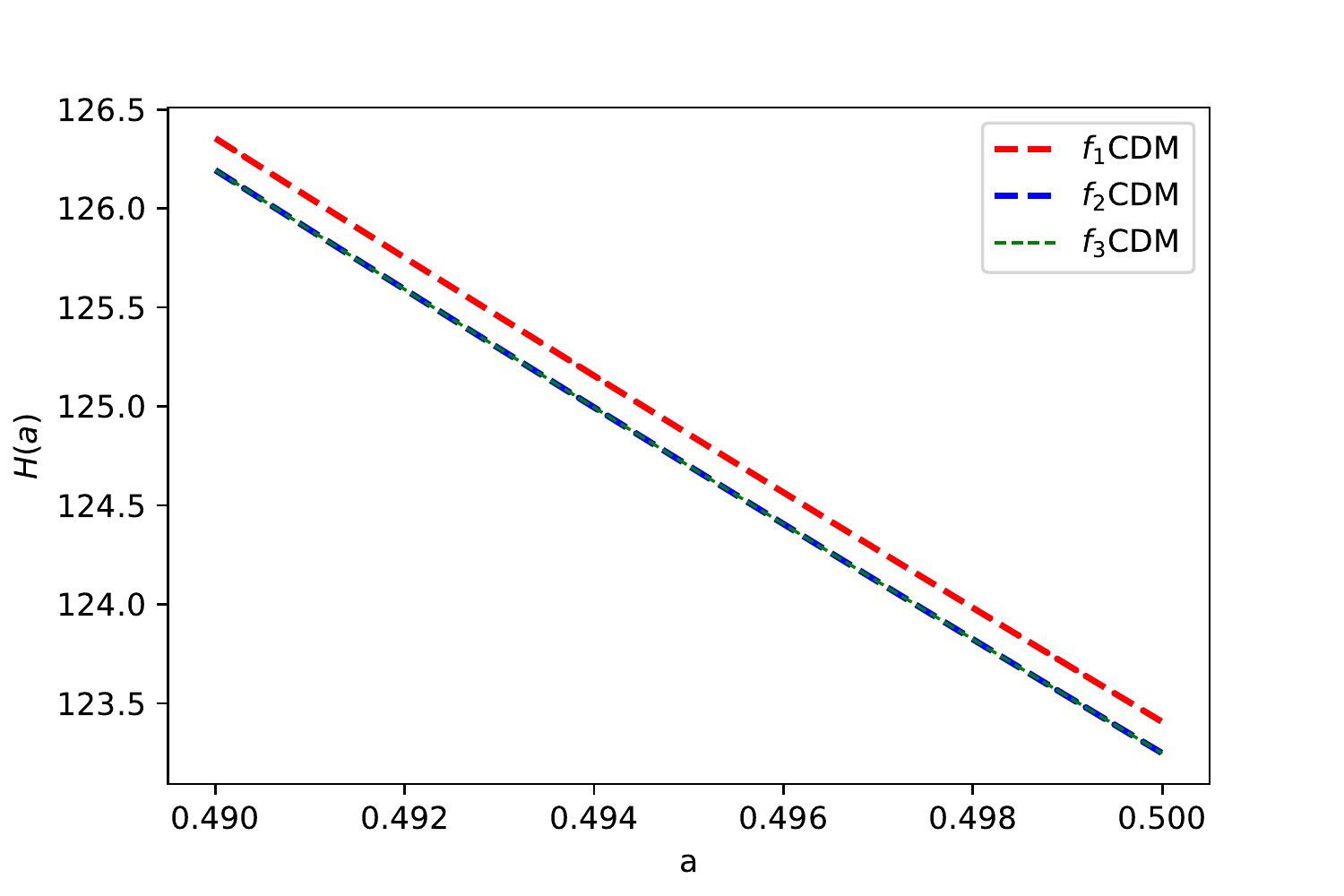}
\caption{Top: beaviour of EoS, $\omega_T$, of Eq.(\ref{eq:omegaeff}) for the $f_{1}$CDM (red), $f_{2}$CDM (blue) and $f_{3}$CDM (green) models assuming $\Omega_T = 0.7$ and $H_0=70$ and also the model free parameter at values $0.1$ (solid line, left panel) and $0.01$ (dashed line, right panel). Note the different scale between the two plots.
Middle: Background evolution for the three models for the same choice of colors and values above, in the left panel considering f(T) parameter at values $0.1$ and in the right panel for $0.01$.
Bottom: Background evolution for the three models in the scale factor range [0.49 - 0.5]
}
\label{fig:w_Ha}
\end{figure*}
%
%
\section{Specific $f(T)$ models}
\label{Sec:fT_models}

We choose to analyze three $f(T)$ functions well known in literature for being viable models, i.e. passing the basic observational tests \cite{Nunes:2016qyp,Capozziello:2017bxm,Nunes:2018xbm}. 
{Noteworthy, $f(T)$ models are also able  to describe the inflationary evolution, as  discussed in \cite{Oikonomou:2017isf}. In particular, they are capable of  achieving the smoothing during the inflationary era and ensuring the linear perturbation up to the BBN epoch.} 
Here, we introduce their forms and derive their background evolution. Also, we show the theoretical observational predictions of temperature anisotropy power spectrum and the EE-mode correlation spectrum for each of them, and we summarize the current state of the art of the constraint of their free parameters.

 \begin{figure*}
\centering
\includegraphics[width=0.9\textwidth]{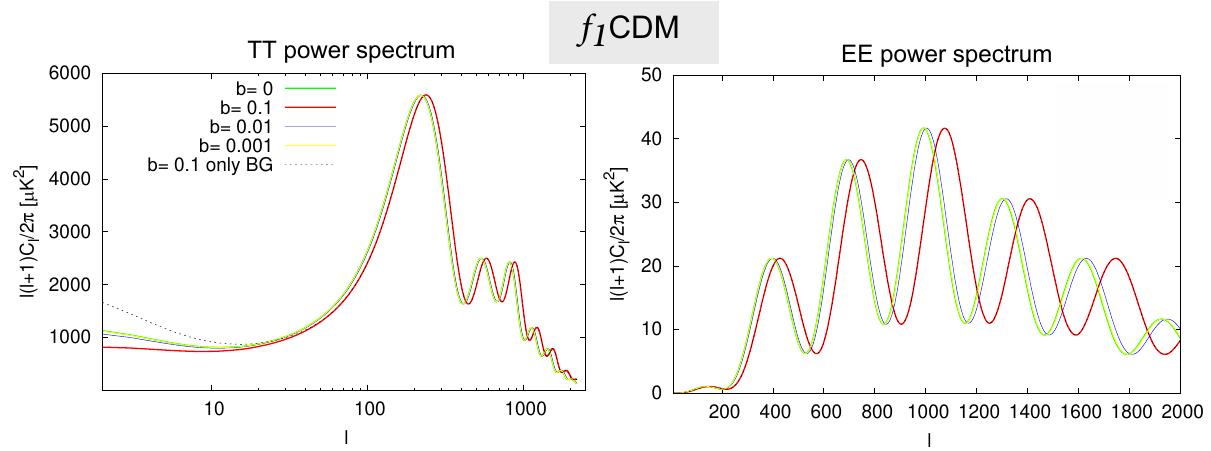}
\includegraphics[width=0.9\textwidth]{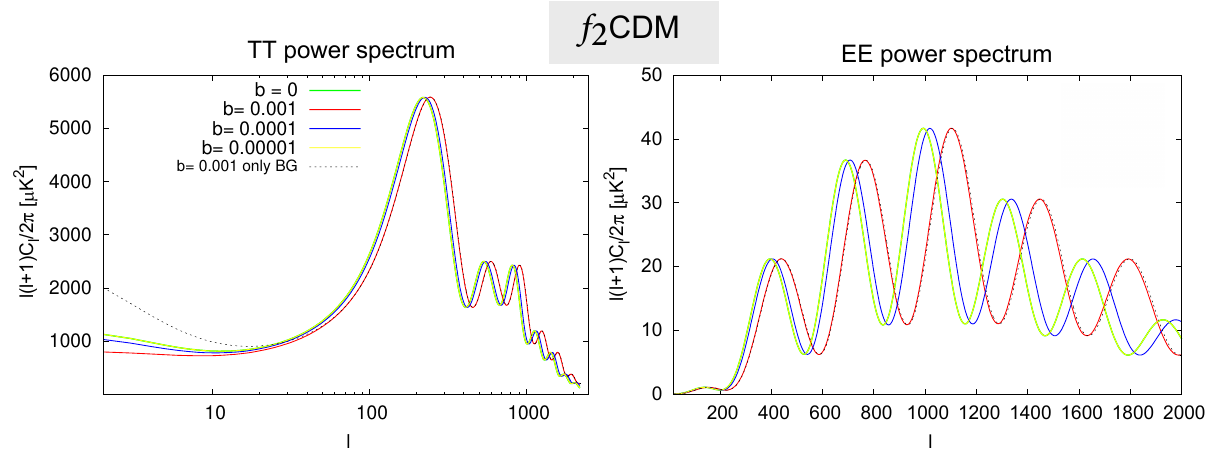}
\includegraphics[width=0.9\textwidth]{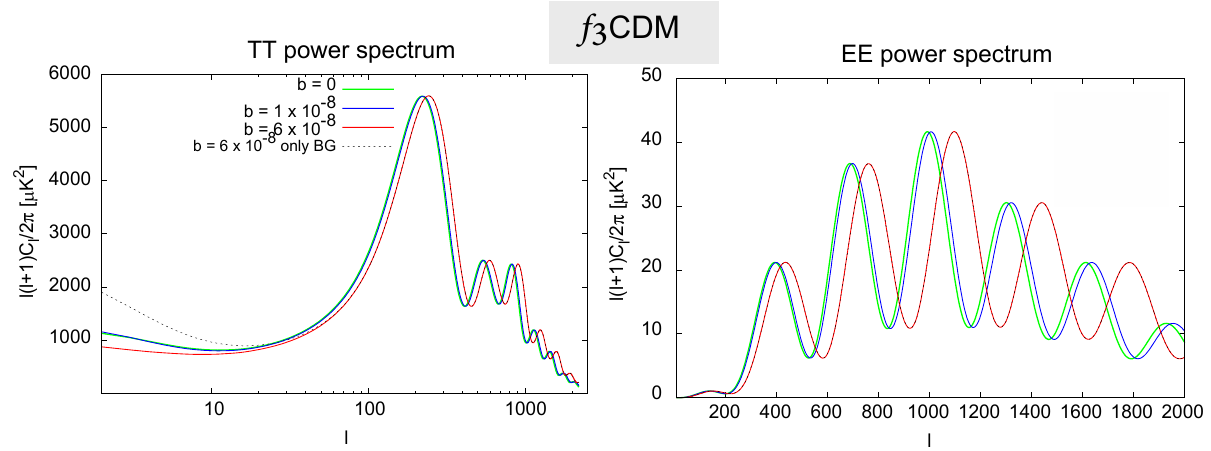}
\caption{ TT and EE correlation CMB anisotropy power spectra for $f_1$CDM (top panels), $f_2$CDM (middle panels) and $f_3$CDM (bottom panels), using several values of the $f(T)$ free parameter, $b$. For each model we draw, with dotted line, the case where only the Background (BG) evolution is considered, i.e. the linear perturbation evolution has not been included. By showing the two cases we want to underline the effect of considering  primordial perturbations in the analysis of the models. In  literature, only background equations are considered when there is a comparison of the theory with the data, and this is mainly due to both  the specific interest of testing whether these models can drive the current acceleration of the Universe and to the difficulty of considering perturbative evolutions in the simulation codes. Here, we show the two observational predictions (with and without perturbations)} . 
\label{fig:power_spectra1}
\end{figure*}
\begin{figure*}
\centering
\label{fig:2D_Yp-Omb}
\includegraphics[width=0.45\textwidth]{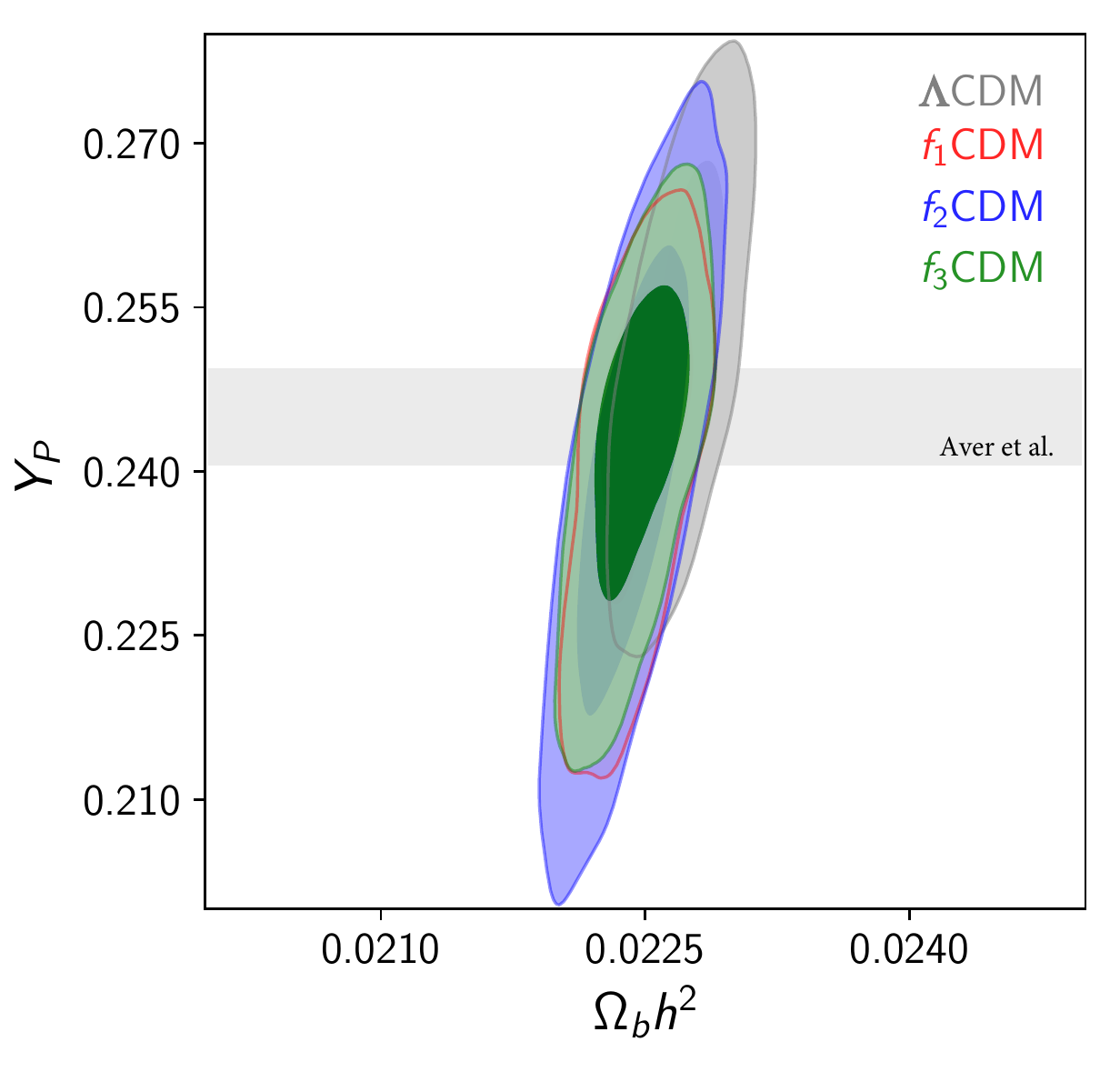}
\includegraphics[width=0.43\textwidth]{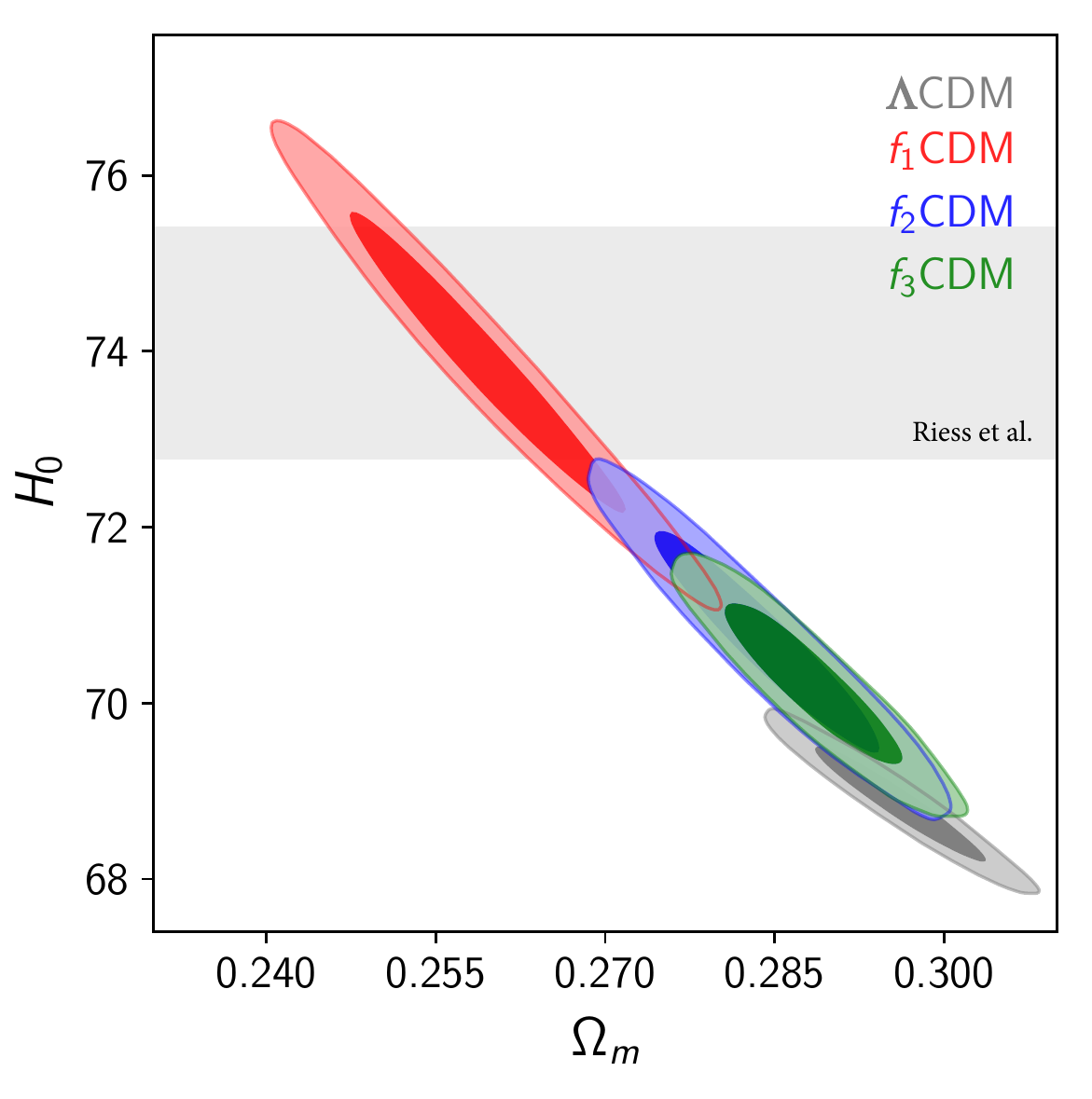}
\caption{ Left: $Y_p$-$\Omega_b$ plane for our analysis with free $Y_p$ models. The grey region indicate the $1\sigma$ estimation of Aver et al. $Y_p = 0.2449 \pm 0.0040$;
Right: $H_0$-$\Omega_m$ plane for our analysis with free $Y_p$ models. The grey region indicate the latest constrain on Hubble constant of Riess (2019), $H_0$ = $74.03 \pm 1.42$ km/s/Mpc .
}

\end{figure*}
%
\begin{itemize}
\item
The first scenario is the power-law model (hereafter $f_{1}$CDM) with
\begin{equation}
\label{fT_powerlaw}
f (T)   = \beta \left(-T \right)^{b} ,\
\end{equation}
 that recovers the GR form, $T+ f(T) = T -2\Lambda$, for $b=0$ and $\Lambda =- \beta / 2$ \cite{Bengochea:2008gz}. Substituting this $f(T)$ form into the first Friedmann equation at present epoch, we obtain the relation between the two parameters, $ \beta=(6H_0^2)^{1-b}\frac{\Omega_{T0}}{2b-1}$, with $\Omega_{T0} = 1-\Omega_{m0} - \Omega_{r0}$.

The $y_T(a,b)$, in the background evolution of Eq.(\ref{eq:Ha}), reads as $y_T(a,b)= \Omega_{T0} E(a)^{2b}$ \cite{Nesseris:2013jea}
that reduces to $\Lambda$CDM cosmology for $b= 0$, while, for  $b= 1/2$,  it gives rise to the Dvali-Gabadadze-Porrati (DGP) model \cite{Dvali:2000hr}. At the same time, we can write the EoS  Eq.(\ref{eq:omegaeff}) as

\begin{equation}\label{eq:omegaeff_powerlaw}
    \omega_{T} = \frac{b-1}{1-b \Omega_{T0} E(a)^{2(b-1)}}
    \end{equation}

that reduces to a constant value $\omega_{T}=-1$ for $b=0$. Its behaviour is shown in Fig.(\ref{fig:w_Ha}), top panels, with red lines. In particular, we assume the values $b=0.1$, solid line, and $b=0.01$, dashed line. We can see that, in both cases, the today EoS value  converges to values close to  $\omega_{T0} = -1$, and depending on $b$, it can assume slightly (negligible) higher values, with a variation up to $ 10^{-1}$ at small scales. We also note that $f_{1}$CDM is the model with the behavior that most differs from the others, both in $\omega_{T}$ and in $H(a)$ evolution.

Previous results show that, using only BBN data, is possible to put an upper-limit as $b < 0.94$ \cite{Capozziello:2017bxm}, while using large scale data, it  is possible to constrain $b=0.033^{+0.043}_{-0.035}$ by Cosmic Chronometers (CC), and $b=0.051^{+0.025}_{-0.019}$ when also SNe Ia and Baryonic Acoustic Oscillations (BAO) are considered \cite{Nunes:2016qyp}.
The model was tested also using measurements from quasar absorption lines and radio quasars \cite{Qi:2017xzl}, and also several large scale data combinations \cite{Anagnostopoulos:2019miu,Xu:2018npu}. Finally, including CMB by Planck (2015) data, joined with BAO and $H_0$ measurements, the most stringent constraint is obtained, $b=0.005 \pm 0.002$ \cite{Nunes:2018xbm}.

Using the approach described in the previous Section, we draw the theoretical observational predictions of the temperature anisotropy and the EE correlation power spectra for this model in the top panel of Fig.(\ref{fig:power_spectra1}), assuming several values for the free $f(T)$ parameter. We can see that higher $b$ means a shift of the spectra to higher multipoles, which implies, among other things, a degeneration of this parameter with the curvature of the Universe and the current  expansion. Note that our observational predictions are fully in agreement with the ones obtained in Newtonian gauge choice \cite{Nunes:2018xbm}. Furthermore, we draw (dotted line) the case without considering the evolution of linear perturbations, i.e. calculating only the background evolution, and we see that the power at low multipoles of TT spectra is particularly affected.

It is important to remark that the $f(T)$ power law models are  selected by the existence of  Noether symmetries as shown in  \cite{Basilakos:2013rua}. This result can be seen as a  criterion to select physical models \cite{Capozziello:1996bi,Paliathanasis:2012at} which allows to reduce and, eventually, integrate  the  equations of motion. In particular, it is possible to find exact cosmological solutions for the form $f_0 T^n$ which  lead to the background evolution of the $\rho_T$ density as $a(t)=a_{0}t^{2b/3}$ and $H(t)=\frac{\dot{a}}{a}=\frac{2b}{3t}$.
In this case, the background evolution for the total density reads as:
\begin{equation}
\label{eq:HE_tot}
H(a)^2 = H_0^2 \left[ \Omega_{m0} a^{-3} + \Omega_{r0} a^{-4} + \Omega_{T0} a^{-3 b} \right]
\end{equation}
Let us stress that this Noether solution is calculated for an action of the form $ {\cal A}_T = \frac{1}{16\pi G}\int{d^4xe f(T)}$.
\item
The second scenario is the square-root-exponential (hereafter $f_{2}$CDM) also called {\textit{Linder model}} \cite{Linder:2010py}, with
\begin{equation}
f(T)=\alpha T_{0}(1-e^{-p\sqrt{T/T_{0}}})
\end{equation}
where the relation between the two parameters is $\alpha=\frac{\Omega_{T0}}{1-(1+p)e^{-p}}$. The first Friedmann equation leads to $y(a, p)= \Omega_{T0}\frac{\left[ 1- \left(1+ p E(a)\right) e^{-pE(a)} \right]}{1-(1+p)e^{-p}}$ \cite{Nesseris:2013jea}, that reduces to $\Lambda$CDM cosmology for $p\rightarrow+\infty$. The EoS for this model is

\begin{equation}
\label{eq:omegaeff_linder}
    \omega_{T} = -\frac{ e^{p E(a)} (e^p-1-p) \left[ 2 ( e^{p E(a)} - 1) - p E(a) ( pE(a) + 2) \right]}{(e^{p E(a)}-1 - p E(a)) \left[ 2 e^{p E(a)} (e^p - 1 -p)  + \Omega_{T0} e^p p^2 \right]}
\end{equation}

and it  is drawn with blue lines in Fig.(\ref{fig:w_Ha}).
Generally, instead of the parameter $p$, its inverse is used, that is  $b \equiv 1/p$. This is because the limit $p\rightarrow+\infty$ is equivalent to $b \equiv 1/p \rightarrow 0^{+}$, and the latter limit is considered more proper to be treated in the analyses. Previous works constrained $b=0.111^{+0.035}_{-0.110}$, using CC data, while the joint analysis $CC+SNeIa+BAO$ allows for $b=0.132^{+0.043}_{-0.130}$ \cite{Nunes:2016qyp} while only BBN data cannot impose constraints on the parameter value \cite{Capozziello:2017bxm}. As in the previous case, we see that the use of the CMB likelihood can significantly increase the precision on the constraint of $b$: in this case, an order of magnitude of $10^{-5}$ is expected.

We show the prediction for the TT and EE spectra in middle panels of Fig.(\ref{fig:power_spectra1}), assuming several values for the free parameter $b$. Also in this case, we can see a shift of the spectra to higher multipoles for increasing values of $b$.
\item
The last scenario we analyze is the the exponential form (hereafter $f_3$CDM) \cite{Linder:2009jz,Bamba:2010wb}

\begin{equation}
 f(T)=\alpha T_{0}(1-e^{-pT/T_{0}})
\end{equation}

 with $\alpha=\frac{\Omega_{m0}}{1-(1+2p)e^{-p}}$. The background evolution can be written as \cite{Nesseris:2013jea}
 \[
 y(a,p)= \Omega_{T0} \frac{1}{1-(1+2p)e^{-p}} \left[ 1- \left(1+ {2pE(a)^2}\right) e^{-pE(a)^2} \right]\,,
 \]
 which  reduces to $\Lambda$CDM cosmology for $p\rightarrow+\infty$ (or $b \equiv 1/p \rightarrow 0^{+}$). At the same time, the EoS reads as

\begin{equation}\label{eq:omegaeff_expo}
\scalebox{0.9}{
$    \omega_{T} = -\frac{ e^{p E(a)^2} (e^{p}-1-2p) \left[e^{p E(a)^2}-1- p E(a)^2 (1 + 2 p E(a)^2) \right]} {(e^{p E(a)^2} -1 - 2p E(a)^2) \left[e^{p E(a)^2} ((e^{p}-1 - 2p) - p \Omega_{T0} e^{p}  (1 - 2p E(a)^2)\right]}$}
\end{equation}

and it is plotted in Fig.(\ref{fig:w_Ha}) with green lines. We note that $f_2$CDM and $f_3$CDM models show similar behaviours, indicating that the presence of the root in exponent of the exponential function does not give any observable difference in the $H(a)$ evolution, for the range of values we are considering. At the same time, looking for the TT and EE spectra predictions (shown in bottom panels of Fig.(\ref{fig:power_spectra1})) we note that a precision of $10^{-8}$ is required on the $b$ parameter to describe the observations, unlike in case $f_2$CDM.

The current bounds on this model constrain $b=0.106^{+0.052}_{-0.090}$ using $CC$ data, while the joint analysis $CC+SNeIa+BAO$ allows for $b=0.090^{+0.041}_{-0.080}$ \cite{Nunes:2016qyp}, and also in this case, the BBN data cannot impose constraints on the parameter value \cite{Capozziello:2017bxm}. These estimates are far from what is required by the TT spectrum to describe CMB observations, we can infer that analysis with the full CMB likelihood can significantly improve the constraint on this model.

 \end{itemize}
\begin{figure*}[t!]
\centering
\includegraphics[width=0.32\textwidth]{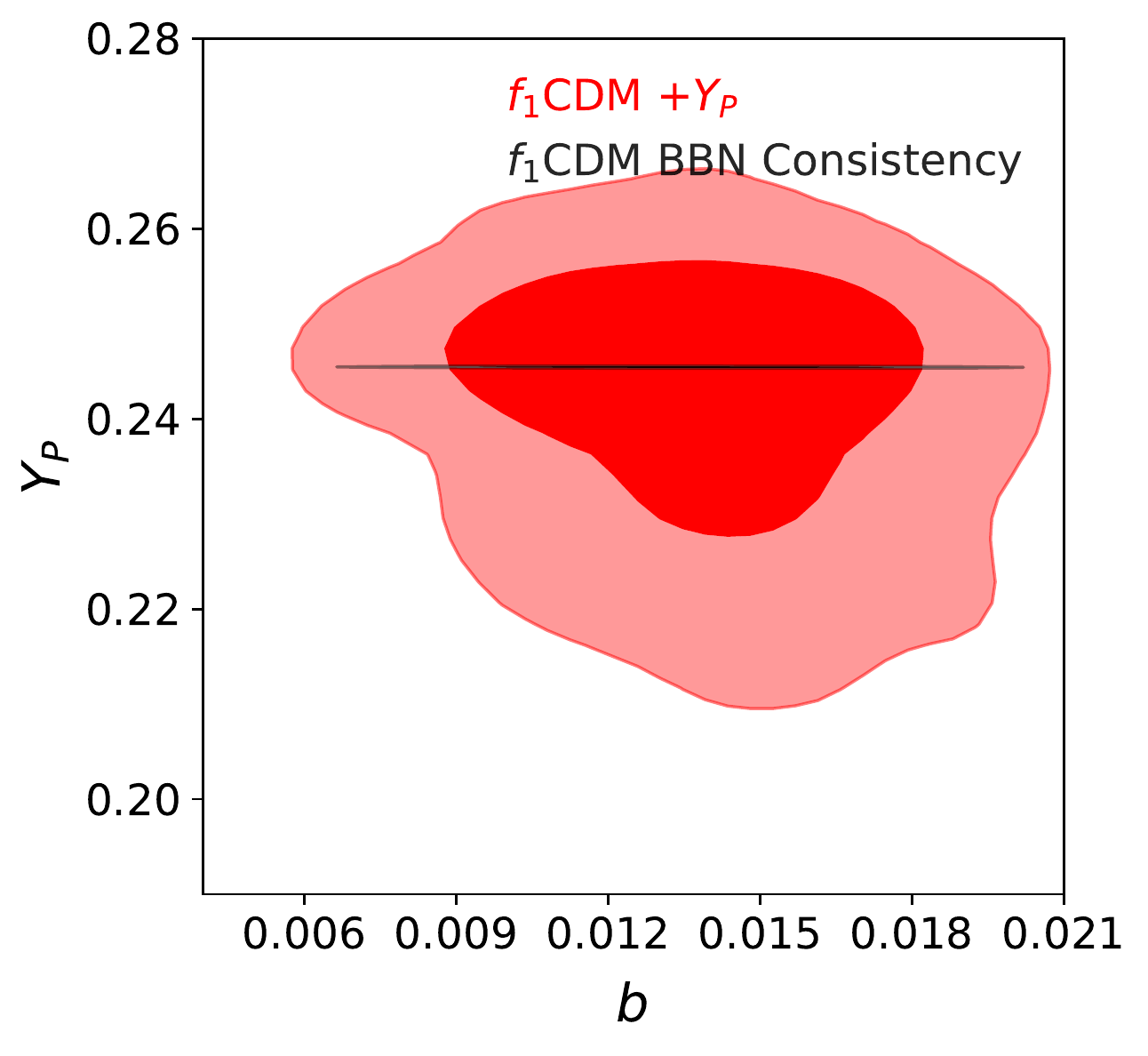}
\includegraphics[width=0.3\textwidth]{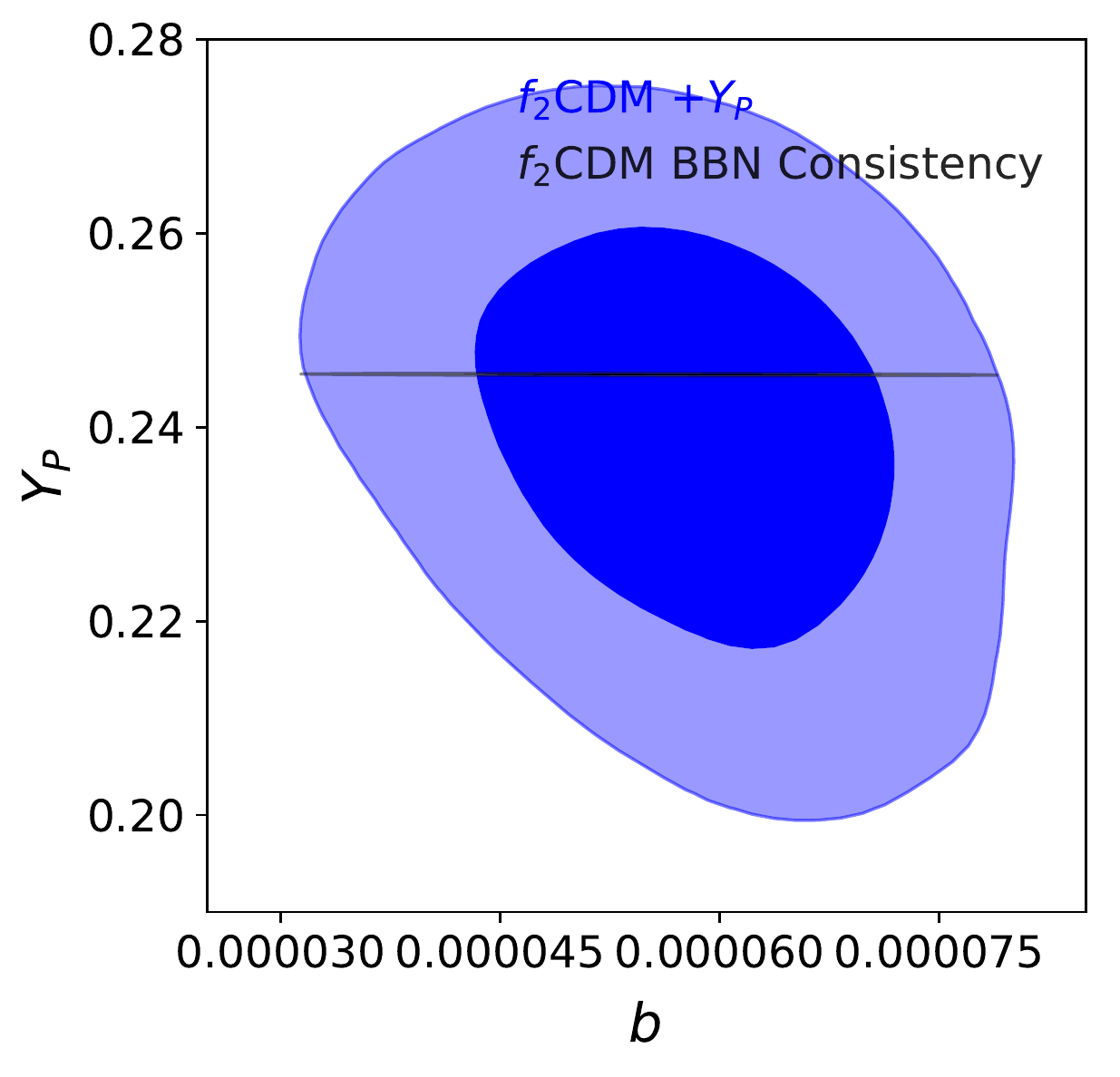}
\includegraphics[width=0.3\textwidth]{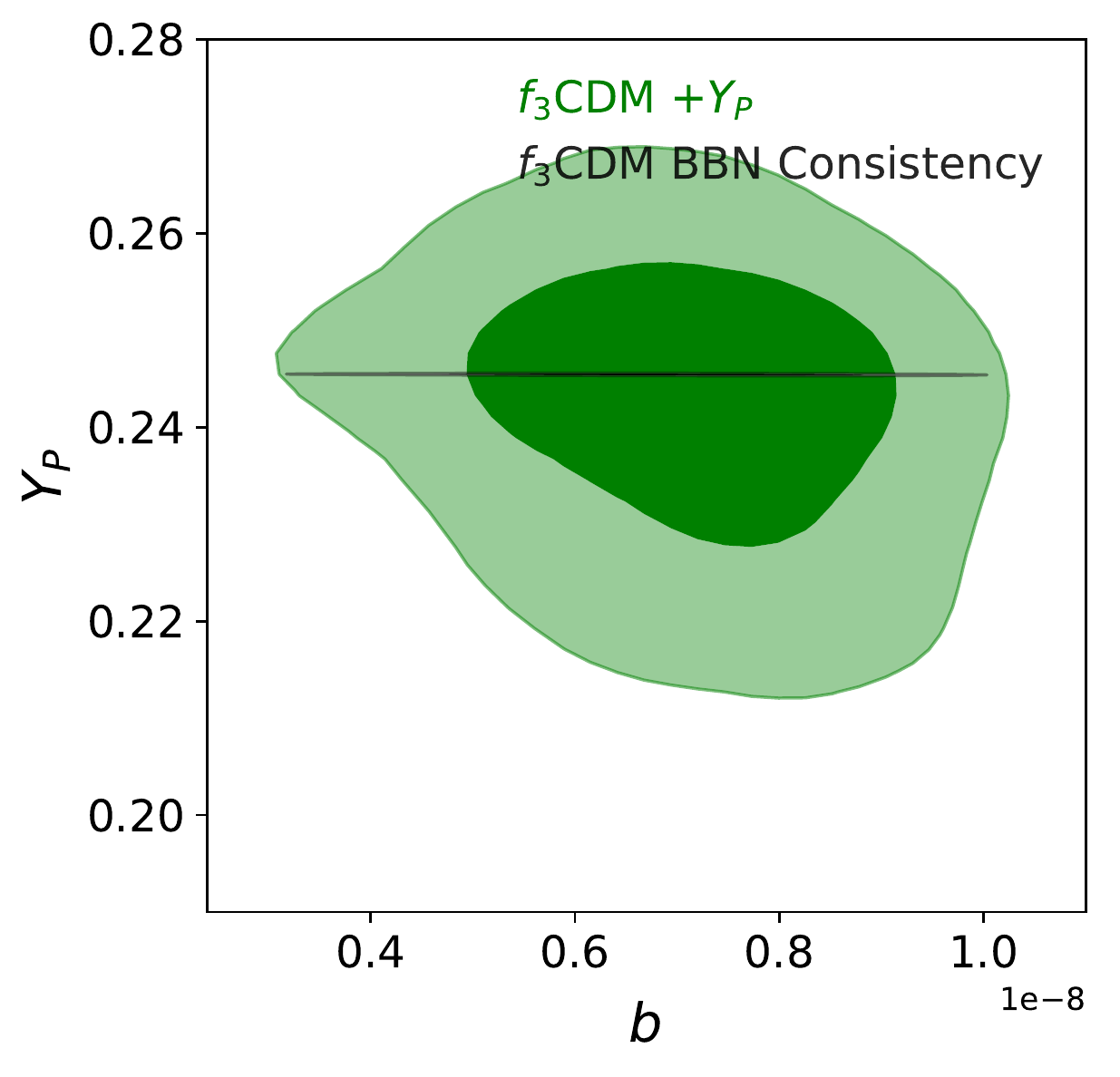}
\caption{ $Y_p$-$b$ plane for our analysis. Left: $f_1$CDM model;  Middle: $f_2$CDM model, Right: $f_3$CDM model.}
\label{fig:2D_Yp-b}
\end{figure*}

Let us now  modify the CosmoMC package \cite{Lewis:2002ah} to include  the background and perturbations evolution for each model, and perform a Monte Carlo Markov chain exploration of the parameters space.

\section{Analysis Method}
\label{Sec:Analysis}
In our analysis, we consider the minimal $\Lambda$CDM model as the reference model, with the usual set of cosmological parameters: the baryon density, $\Omega_bh^2$, the cold dark matter density, $\Omega_ch^2$, the ratio between the sound horizon and the angular diameter distance at decoupling, $\theta$, the optical depth, $\tau$, the primordial scalar amplitude, $A_s$, and the primordial spectral index $n_s$. For each $f(T)$ model, we consider also one more free parameter given by the specific $f(T)$ form, by modifying the CAMB code to reflect the models described in the previous Section.

Also, we consider both the cases where the BBN consistency is considered or not.
In the first case, the primordial helium fraction value is derived from the BBN consistency relation as a function of the baryon and radiation densities, and we use the PArthENoPE fitting table\footnote{PArthENoPE website: http://parthenope.na.infn.it/} to calculate such a primordial abundances of helium and deuterium. We refer to this case with ``$f_i$CDM BBN Consistency". Instead, in the second case, the helium fraction is considered as a free parameter of the model, and we refer to this case with ``$f_i$CDM $+Y_p$". This choice to treat $Y_p$ as a model parameter, and not derived from the BBN consistency relation, has been recently explored in the literature  to resolve the so-called $H_0$ tension \cite{DiValentino:2014cta,Schoneberg:2019wmt,Bernal:2016gxb,Benetti:2017gvm}. Indeed, an higher radiation energy density, i.e. an higher $Y_p$ value, imply a larger expansion rate of the Universe \cite{Hamann:2007sb,Iocco:2008va}. In this work, we choose to explore a free $Y_p$ to study if $f(T)$ gravity spontaneously recovers the primordial abundances predicted by the theory, and if the free parameter of $f(T)$ models shows any degeneration with the BBN abundance.
It is worth mentioning that other BBN codes are available and may give slightly different values of primordial abundances, however with an error inside $\Delta{ }Y_p = 0.0003$ \cite{Cyburt:2015mya}. Here we use the code most widely employed and  adopted also  by the Planck collaboration 
\cite{Aghanim:2018eyx}.

{Another important remark is in order at this point.
 In our analysis we are assuming that, at the end of BBN era, the Universe evolves with temperatures lower than 0.1 MeV.  In other words,  the initial conditions for the following Universe evolution, such as linear cosmological perturbations and non-interacting Cold Dark Matter,  are in agreement with the $f(T)$ models we are considering. In this context, it is worth mentioning that, in the framework of $f(T)$ gravity, it is possible to address the today observed accelerated expansion as well as the clustering phenomena as effects related to the torsion fluid. Indeed, $f(T)$ gravity can mimic dark matter effects  contributing to the structure formation, i.e. corrections to the Newtonian potential induced by $f(T)$ models can successfully address dark matter issues at any scale (see e.g. \cite{Finch:2018gkh} for a detailed discussion of dark matter through $f(T)$ gravity in galaxies). According to these results,  the BBN codes, based on CDM, can be assumed unchanged in $f(T)$ framework. As a consequence, because $f(T)$ gravity can mimic dark matter, all the initial conditions for the post-BBN phase result the same. Vice versa, viable $f(T)$ models can successfully satisfy the BBN constraints. See  \cite{Capozziello:2017bxm} for a discussion on this point}.

In our analysis,  we choose to work with flat priors, and consider purely adiabatic initial conditions, fixing the sum of neutrino masses to $0.06~eV$. In particular, for the helium fraction $Y_p$,  we explore the prior [0.1 : 0.6].
At this point, it is worth noticing that we can also test logarithmic priors for the new parameters, which in principle could  be more suitable  for the $f_2$ and $f_3$ models. Nevertheless   we do not appreciate any significant improvement in the estimation and then we  decide to show the results obtained only for the flat-prior choice because this procedure  allow a more immediate comparison of the results.

We consider the joint data set of the following measurements:

\begin{table*}
\centering
\caption{
$68\%$ confidence limits for the $f(T)$CDM and $\Lambda$CDM analysis using CMB+lensing+BAO+R19+Pth+DES data}
\label{tab:Parameters}
\scalebox{1}{
\begin{tabular}{|c|c|c|c|c|}
\hline
\multicolumn{5}{|c|}{BBN Consistency}\\
\hline
\multicolumn{1}{|c|}{ }&
\multicolumn{1}{c|}{ $\Lambda$CDM }&
\multicolumn{1}{c|}{ $f_1$CDM}&
\multicolumn{1}{c|}{ $f_2$CDM}&
\multicolumn{1}{c|}{ $f_3$CDM}\\
\hline
$100\,\Omega_b h^2$ 	
& $2.264 \pm 0.013$
& $2.251 \pm 0.013$
& $2.248 \pm 0.014$
& $2.249 \pm 0.014$
\\
$\Omega_{c} h^2$	
& $0.1170 \pm 0.0008$
& $0.1183 \pm 0.0008$
& $0.1189 \pm 0.0010$
& $0.1189 \pm 0.0011$
\\
$\tau$
& $0.061 \pm 0.008$
& $0.056 \pm 0.007$
& $0.054 \pm 0.007$
& $0.054 \pm 0.008$
\\
${\rm{ln}}(10^{10} A_s)$
& $ 3.053 \pm 0.015 $
& $ 3.045 \pm 0.014 $
& $ 3.041 \pm 0.014 $
& $ 3.041 \pm 0.015 $
\\
$n_s$
& $ 0.972 \pm 0.004$
& $ 0.968 \pm 0.004$
& $ 0.967 \pm 0.004$
& $ 0.967 \pm 0.004$
\\
$b$
& $ - $
& $ (1.4 \pm 0.3 ) \times 10^{-2}$
& $ (5.6 \pm 0.9 ) \times 10^{-5}$
& $ (6.8 \pm 1.3 ) \times 10^{-9}$
\\
$Y_p$ \footnote{Derived parameter obtained from BBN consistency.}
& $ 0.24550 \pm 0.00005$
& $ 0.24545 \pm 0.00005$
& $ 0.24543 \pm 0.00005$
& $ 0.24544 \pm 0.00005$
\\
$H_0$
& $ 68.79 \pm 0.36$
& $ 73.85 \pm 1.05$
& $ 70.67 \pm 0.82$
& $ 70.14 \pm 0.62$
\\
$\sigma_8$
& $ 0.806 \pm 0.006 $
& $ 0.850 \pm 0.010 $
& $ 0.824 \pm 0.009 $
& $ 0.818 \pm 0.007 $
\\
\hline
$\Delta DIC$ 	
& -
& \text{strongly preferred}
& \text{moderately preferred}
& \text{moderately preferred}
\\
\hline
\multicolumn{5}{|c|}{$Y_p$ free }\\
\hline
\multicolumn{1}{|c|}{ }&
\multicolumn{1}{c|}{ $\Lambda$CDM+$Y_p$ }&
\multicolumn{1}{c|}{ $f_1$CDM+$Y_p$}&
\multicolumn{1}{c|}{ $f_2$CDM+$Y_p$}&
\multicolumn{1}{c|}{ $f_3$CDM+$Y_p$}\\
\hline
$100\,\Omega_b h^2$ 	
& $2.270 \pm 0.017$
& $2.247 \pm 0.016$
& $2.242 \pm 0.020$
& $2.250 \pm 0.017$
\\
$\Omega_{c} h^2$	
& $0.1170 \pm 0.0008$
& $0.1184 \pm 0.0008$
& $0.1190 \pm 0.0010$
& $0.1190 \pm 0.0010$
\\
$\tau$
& $0.062 \pm 0.008$
& $0.056 \pm 0.007$
& $0.053 \pm 0.007$
& $0.053 \pm 0.007$
\\
${\rm{ln}}(10^{10} A_s)$
& $3.055 \pm 0.016$
& $3.044 \pm 0.014$
& $3.038 \pm 0.015$
& $3.039 \pm 0.015$
\\
$n_s$
& $ 0.974 \pm 0.006$
& $ 0.967 \pm 0.005$
& $ 0.965 \pm 0.007$
& $ 0.966 \pm 0.006$
\\
$b$
& $ - $
& $ (1.4 \pm 0.3 ) \times 10^{-2}$
& $ (5.7 \pm 0.9 ) \times 10^{-5}$
& $ (7.1 \pm 1.3 ) \times 10^{-9}$
\\
$Y_p $ \footnote{Free parameter of the model.}
& $ 0.250 \pm 0.011$
& $ 0.243 \pm 0.010$
& $ 0.239 \pm 0.014$
& $ 0.243 \pm 0.010$
\\
$H_0$
& $ 68.87 \pm 0.41 $
& $ 73.86 \pm 1.09 $
& $ 70.68 \pm 0.79 $
& $ 70.21 \pm 0.59 $
\\
$\sigma_8$
& $ 0.807 \pm 0.007 $
& $ 0.851 \pm 0.011 $
& $ 0.823 \pm 0.009 $
& $ 0.818 \pm 0.008 $
\\
\hline
$\Delta DIC$ 	
& -
& \text{strongly preferred}
& \text{moderately preferred}
& \text{moderately preferred}
\\
\hline
\end{tabular}}
\end{table*}

\begin{itemize}

\item CMB measurements, through the Planck (2018) data~\cite{Aghanim:2019ame}, using Plik ``TT,TE,EE+lowE" likelihood by combination of temperature power spectra and cross correlation TE and EE over the range $\ell \in [30, 2508]$, the
low-$\ell$ temperature Commander likelihood, and the low-$\ell$ SimAll EE likelihood. We refer to this data set as ``CMB";

\item The lensing reconstruction power spectrum from the latest Planck satellite data release (2018)~\cite{Aghanim:2019ame,Aghanim:2018oex}, hereafter indicated with ``lensing";

\item Baryon Acoustic Oscillation (BAO): we use  distance measurements from 6dFGS ~\cite{Beutler:2011hx}, SDSS-MGS~\cite{Ross:2014qpa}, and BOSS DR12~\cite{Alam:2016hwk} surveys, as considered by the Planck collaboration;

\item Hubble constant of latest Riess (2019) work (R19), $H_0$ = $74.03 \pm 1.42$ km/s/Mpc ~\cite{Riess:2019cxk}, that is in tension at 4.4$\sigma$ with CMB estimation within the minimal cosmological model. This measurement is implemented by default in the package {\sc CosmoMC} by imposing a Gaussian prior for the Hubble parameter constraint.

\item Pantheon compilation~\cite{Scolnic:2017caz} of 1048 SNe Ia in the redshift range $0.01< z <2.3$, which provides accurate relative luminosity distances, hereafter indicated with ``Pth";

\item  Dark Energy Survey Year-One (DES) results that combine galaxy clustering and weak gravitational lensing measurements, using 1321 square degrees of imaging data~\cite{Abbott:2017wau}.

\end{itemize}

\section{Results and Conclusions}
\label{Sec:Results}
The results of the  analysis  are summarized in Tab. (\ref{tab:Parameters}), where the constraints on free parameters of the theory, and some of the derived ones, are shown. Also, in Fig.(\ref{fig:2D_Yp-Omb}) we show the plane $Y_p - \Omega_bh^2$ with superimposed  direct measurements of $Y_P$ by observations of helium and hydrogen emission lines from metal-poor extragalactic H II regions, combined with estimates of metallicity, $Y_P = 0.2449 \pm 0.0040$ \cite{Aver:2015iza}, consistent with the standard BBN estimate, $Y_P = 0.2477 \pm 0.0029$~\cite{Peimbert:2007vm,Cyburt:2015mya,Steigman:2012ve}.
We show that the considered $f(T)$ models are fully in agreement with direct and indirect measurements, i.e. the BBN result based on the Planck determination \cite{Aghanim:2018eyx}. Noteworthy, the helium fraction parameter shows no correlation with the free parameter of the $f(T)$ gravity, avoiding the introduction of degeneration (see Fig.(\ref{fig:2D_Yp-b})).

We note that the introduction of the CMB likelihood in the analysis significantly improves the constraints on the $f(T)$ free model parameter, achieving an accuracy of
$10^{-2}$, $10^{-5}$ and $10^{-9}$ respectively for the case of power law, exponential  and  the  square-root exponential $f(T)$ gravity. Our results confirm previous analysis using the full CMB likelihood \cite{Nunes:2018xbm}, and constrain the $f(T)$ gravity parameter as different from zero at more than 3$\sigma$. This is particularly significant in the light of large scale data analysis results, where $f(T)$ parameters were compatible with zero in 1$\sigma$ \cite{Nunes:2016qyp,Qi:2017xzl,Anagnostopoulos:2019miu,Xu:2018npu,Escamilla-Rivera:2019ulu}. In other words, these results show a preference of the analysed dataset for a deviation from the standard $\Lambda$CDM.
We can infer that cosmic dynamics could constitute a probe for deviation with respect to GR (or TEGR). In particular,
we note that $f_i$CDM models prefers higher $H_0$ values with respect to the $\Lambda$CDM one (see Fig.\ref{fig:2D_Yp-Omb}) 
and using the complete CMB likelihood improves the precision on the constraint of the $f(T)$ parameter and further relaxes the $H_0$ tension, even solving it in the case of $f_1$CDM model. Noteworthy, this occurs both when BBN consistency is considered and when $Y_p$ is treated as a free parameter. That is, a faster expansion is not achieved at the cost of extra amount of primordial abundances or a higher radiation density, but with a modification of gravity.

\begin{figure*}[t!]
\centering
\includegraphics[width=0.8\textwidth]{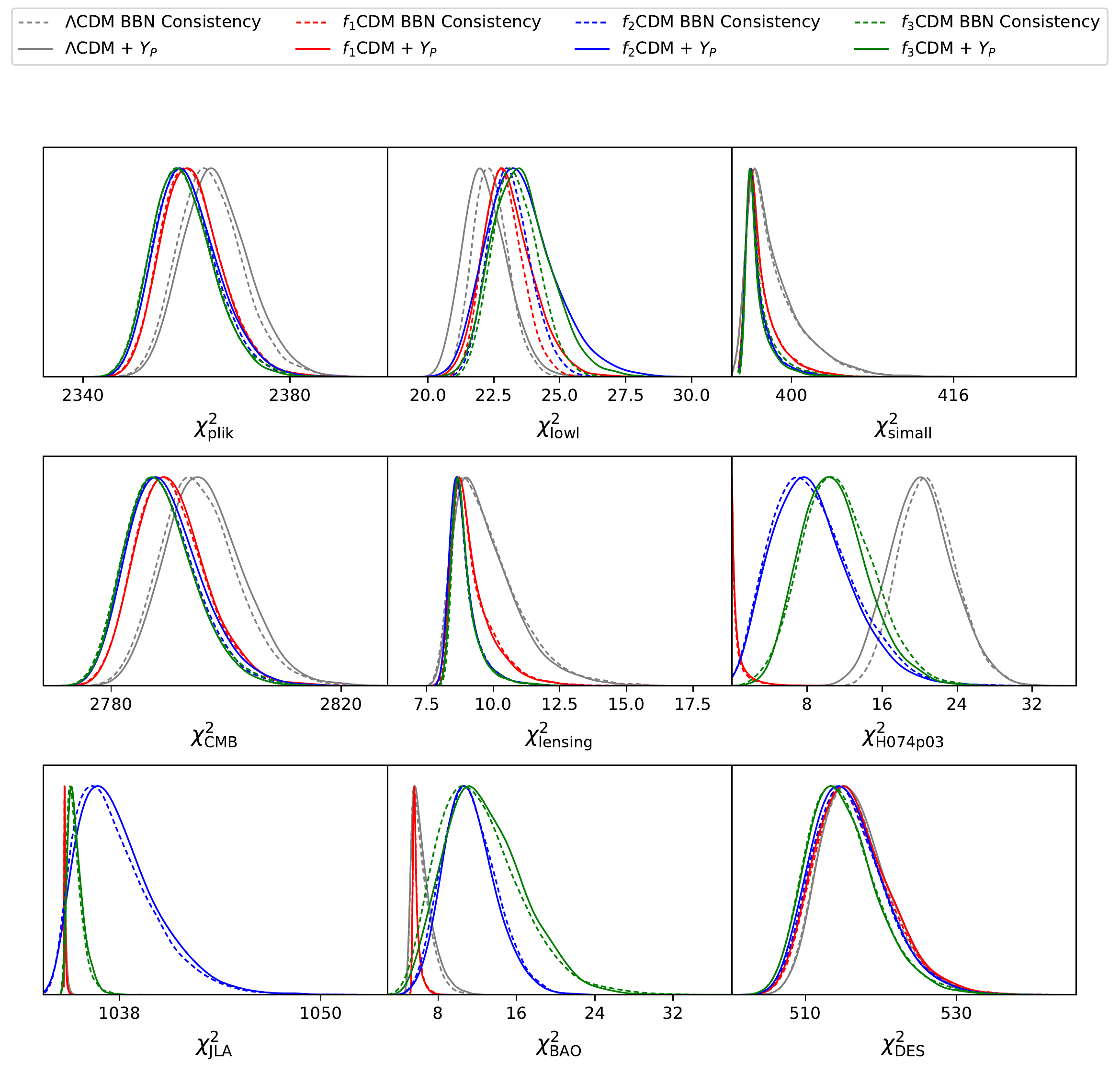}
\caption{Posterior distribution density of the $\chi^2$ values of the several data used in the analysis.}
\label{fig:chi2}
\end{figure*}
Finally, to compare  $f(T)$ models with the $\Lambda$CDM, when constrained with data, we use the Deviance Information Criterion (DIC)~\cite{Spiegelhalter:2002yvw}:
\be
\text{DIC}:= \chi_\text{eff}^2 + 2 p_\text{D},
\ee
where $\chi_\text{eff}^2$ is the effective $\chi^2$ corresponding to the maximum likelihood and $p_\text{D} = \overline{\chi}_\text{eff}^2 - \chi_\text{eff}^2$.  The bar stands for the average of the posterior distribution. The DIC accounts for both the accuracy of fit and the bayesian complexity of the model. In Tab.\ref{tab:Parameters},  we indicate the
\be
\Delta \text{DIC} = \text{DIC}_\text{f(T)} - \text{DIC}_\text{$\Lambda$CDM},
\ee
where we consider the convention based on Jeffreys' scale \cite{stat1,stat2} for which  $\Delta \text{DIC}>10/6/2$ provides, respectively, strong/moderate/weak evidence against  $f(T)$ models.

We find that the analised $f(T)$ models are always preferred over the standard one. This result is to be read as a preference of the data, especially of R19 Gaussian prior, for models with a current scale-dependent evolution.
We detail, in Fig.(\ref{fig:chi2}),  the $\chi^2$ density posterior distributions of each dataset considered, which allows us to understand why the $f_1$CDM model is preferred over others.
Indeed, $f_1$CDM minimizes the $\chi^2_{R19}$, i.e it is more in agreement with the estimate of R19, as it can also be seen in Fig.(\ref{fig:2D_Yp-Omb}).
Also, we note that the high-$\ell$ CMB likelihood, the $\chi^2_{plik}$, also shows lower values in the case of  $f(T)$ models compared to the $\Lambda$CDM. The combination of these two effects brings a $\chi^2$ value about 25 points lower than the standard model for the $f_1$CDM. It is clear that the result would be different if the prior of R19 was removed in the choice of the dataset, since there would be no difference in $\chi^2_{R19}$ shown in Fig.(\ref{fig:chi2}). Furthermore, we would expect different evidences if instead of the $\Lambda$CDM model, we consider more general models like the $w$CDM model as reference, with $w$ different from the standard  EoS of $\Lambda$CDM (see
\cite{Capozziello:2019cav}).\\


In conclusion, in this paper we have considered $f(T)$ extensions of teleparallel  gravity intended as corrections to TEGR where only the torsion scalar $T$ is considered. In particular, we studied power law and exponential  corrections, where the standard $\Lambda$CDM can be easily recovered.
 Specifically, we draw both the background and the linear perturbation evolution for three $f(T)$ models, implementing in a Boltzman solver code the theory and studying the theoretical predictions in the light of both large and small scale data.
Our analysis constrain the free parameters of the theory with unprecedented precision, noting that the recovery of GR is out of more than $3\sigma$. Also, when the helium fraction is treated as a free parameter of the models, its constrained value is fully compatible with both direct measurements of primordial abundance and the standard Big Bang Nucleosynthesis estimate, also allowing for a higher $H_0$ value than the standard cosmological model. Noteworthy, this allows to significantly relax the tension on the value of the today observed Hubble constant,
 but also a worsening of the tension on $\sigma_8$ \cite{MacCrann:2014wfa,Battye:2014qga,Benetti:2017juy}
 since the correlation of the two parameters does not seem to be removed from $f(T)$ models.

Future CMB experiments, as  COrE \cite{DiValentino:2016foa}, Stage IV CMB experiment \cite{Abazajian:2019tiv} and SPT-3G \cite{Benson:2014qhw}, will better constrain the primordial abundances \cite{Salvati:2015wxa}. Also Euclid mission \cite{Amendola:2016saw}, combining it with the latest Planck data and with the future COrE mission, clearly will help in breaking the degeneracy between the cosmological parameters, with a significant  reduction of the error on $Y_P$. Finally, Square Kilometre Array (SKA) mission is proved to be a promising tool to test gravity over a large range of scales and redshifts. 


\section*{Acknowledgements}
The authors acknowledge Istituto Nazionale di Fisica Nucleare (INFN), sezione di Napoli, iniziative specifiche QGSKY and MOONLIGHT-2.
We also acknowledge the use of CosmoMC package. This work was developed thanks to the National Observatory (ON) computational support.

\end{document}